\documentclass[lettersize,journal]{IEEEtran}
\usepackage{amsmath,amsfonts}
\usepackage{algorithmic}
\usepackage{algorithm}
\usepackage{array}
\usepackage[caption=false,font=normalsize,labelfont=sf,textfont=sf]{subfig}
\usepackage{textcomp}
\usepackage{url}
\usepackage{verbatim}
\usepackage{graphicx}
\usepackage{multirow}
\usepackage{color}
\usepackage{booktabs}
\usepackage{graphics}
\usepackage{float}
\usepackage{dblfloatfix}
\usepackage{cite}
\usepackage[pagebackref,breaklinks,colorlinks]{hyperref}

\begin{document}

\title{Artificially Generated Visual Scanpath  Improves Multi-label Thoracic Disease  Classification in Chest X-Ray Images}

\author{Ashish~Verma, Aupendu~Kar, Krishnendu~Ghosh, Sobhan~Kanti~Dhara,~\IEEEmembership{Member,~IEEE}, Debashis~Sen,~\IEEEmembership{Senior~Member,~IEEE}
        and~Prabir~Kumar~Biswas,~\IEEEmembership{Senior~Member,~IEEE}
\thanks{Ashish~Verma, Aupendu~Kar, Krishnendu~Ghosh, Debashis~Sen and Prabir~Kumar~Biswas are associated with the Department of Electronics and  Electrical Communication Engineering, Indian Institute of Technology Kharagpur, India-721302.(e-mail: verma.bao@gmail.com, mailtoaupendu@gmail.com, krishnendug6@gmail.com, dsen@ece.iitkgp.ac.in, pkb@ece.iitkgp.ac.in).} 
\thanks{Sobhan~Kanti~Dhara is associated with the Department of Electronics and Communication Engineering, National Institute of Technology Rourkela, India-769008. (e-mail: dhara.sk@gmail.com).}}



\maketitle

\begin{abstract}
Expert radiologists visually scan Chest X-Ray (CXR) images, sequentially fixating on anatomical structures to perform disease diagnosis. An automatic multi-label classifier of diseases in CXR images can benefit by incorporating aspects of the radiologists' approach. Recorded visual scanpaths of radiologists on CXR images can be used for the said purpose. But, such scanpaths are not available for most CXR images, which creates a gap even for modern deep learning based classifiers. This paper proposes to mitigate this gap by generating effective artificial visual scanpaths using a visual scanpath prediction model for CXR images. Further, a multi-class multi-label classifier framework is proposed that uses a generated scanpath and visual image features to classify diseases in CXR images. While the scanpath predictor is based on a recurrent neural network, the multi-label classifier involves a novel iterative sequential model with an attention module. We show that our scanpath predictor generates human-like visual scanpaths. We also demonstrate that the use of artificial visual scanpaths improves multi-class multi-label disease classification results on CXR images. The above observations are made from experiments involving around 0.2 million CXR images from 2 widely-used datasets considering the multi-label classification of 14 pathological findings. Code link: \href{https://github.com/ashishverma03/SDC}{https://github.com/ashishverma03/SDC} 
\end{abstract}


\section{Introduction}
\label{sec:intro}

In radiological imaging, X-ray is the most popular, familiar, and widely acquired imaging modality for diagnosis~\cite{ccalli2021deep, hossain2023non} and other applications~\cite{gao20233dsrnet, xia2021classify}. The importance of Chest X-Ray (CXR) for primary investigation in different thoracic diseases results in a huge volume of CXR acquisition~\cite{raoof2012interpretation, ccalli2021deep}. However, the interpretation of such images is tedious and time-consuming, and requires radiologists’ expertise and experience~\cite{baltruschat2019comparison, ccalli2021deep}. In many world regions, the number of radiologists is in general substantially less than the required, given the huge volume of CXR captured. Moreover, the strenuous nature of the task may cause fatigue-related errors, which may have significant consequences~\cite{jaiswal2019identifying}. Therefore, a computer-aided system is highly required for CXR screening to aid the radiologists, lowering the possibility of mistakes and allowing prioritization of time-sensitive cases~\cite{guan2020multi, ccalli2021deep}. Such a system would do well by focusing on CXR image regions in an order followed by radiologists, who deploy their visual attention specific to the measurement and interpretation of anatomical structures in a CXR.

\textcolor{black}{The emergence of deep learning has prompted widespread exploration of data-driven methodologies for the automated diagnosis of medical conditions utilizing different imaging data modalities \cite{gilakjani2022emotion,gilakjani2023graph,li2024rheumatoid,hassanzadeh2024deep,qin2018computer,azizi2021big}. Among these, chest X-ray (CXR) analysis has gained significant attention. Specifically, with the publication of large labeled CXR image datasets~\cite{wang2017chestx, johnson2019mimic, irvin2019chexpert}, designing automatic screening frameworks for diagnosis has gained popularity~\cite{ccalli2021deep, li2022ragcn}.} 
Different deep models have been designed for various applications like report generation~\cite{yan2021weakly}, 
region segmentation~\cite{chen2022tae}, abnormality detection~\cite{ouyang2020learning}, disease localization~\cite{mansoor2019generic},  and disease classification~\cite{wang2019thorax}. A challenging problem of CXR is to design a single framework that can detect multiple diseases in it, especially in the presence of a high diversity~\cite{ouyang2020learning}. Such a multi-class multi-label classification of diseases in CXR images can benefit by trying to mimic a radiologist's scanning of the image.

Humans perceive an image by scanning it through fast eye movements called saccades between fixations, which results in visual scanpaths. Human visual scanpaths are strongly related to humans' image perception, which has been thoroughly studied \cite{yarbus2013eye, henderson2003human}. Use of visual scanpaths has proven to be effective in different computer vision tasks~\cite{yun2013studying, rosa2015beyond, cheng2023identification} on natural images and videos~\cite{ge2015action, mathe2014actions}.

Radiologists are trained to look into clinically relevant regions of a CXR image and perform the diagnosis. Therefore, the use of radiologists' visual scanpaths on CXR can play a vital role in multi-class multi-label disease classification. Recently, GazeRadar~\cite{bhattacharya2022gazeradar} made a related attempt by performing binary classification in CXR images through radiomics features learned based on recorded fixation locations of radiologists. Discriminative disease features are usually located in small regions of CXR images, which are viewed by radiologists in a systematic manner \cite{kok2016systematic}. The feature viewing pattern usually varies for different diseases~\cite{van2017visual}, which can be learnt from their recorded visual scanpaths.


Recording visual scanpaths of radiologists reviewing CXR images is non-trivial, and thus, it forms a major bottleneck for visual scanpath based automated disease classification in CXR. High-end eye-tracking devices are required for high-precision scanpath recording, which may not be easily available. Therefore, to overcome the bottleneck, it is imperative that a visual scanpath prediction model is learnt, which can generate effective artificial visual scanpaths across different datasets. Such a model can be trained on a CXR image dataset where recorded scanpaths of radiologists are present. 
Very few such datasets are currently available, and REFLACX~\cite{lanfredi2022reflacx}, which is a subset of MIMIC~\cite{johnson2019mimic}, is one such dataset with $2616$ ($\sim1.1\%$ of the whole set) images. An effectively estimated visual scanpath in a CXR image with no recorded scanpath can be leveraged to learn the feature viewing pattern for multi-class and multi-label disease classification.


This paper proposes a framework that first generates a visual scanpath on a CXR image, and then, uses that for multi-class multi-label classification of diseases in that image. 
The overall proposed framework comprises two modules: 
a visual scanpath predictor and a visual scanpath-based multi-class multi-label disease classifier. Our visual scanpath prediction model employs a recurrent neural network that is trained on a limited number of recorded scanpaths of radiologists available for CXR images. Our scanpath predictor is designed to learn the mapping of CXR images' visual features to fixation locations in radiologists' scanpaths along with the sequential dependency among subsequent fixations. Thus, it models the feature viewing pattern of radiologists during our automated classification of diseases. 

For multi-class multi-label disease classification, the scanpath predictor generates a scanpath for the CXR image at hand, and then, the generated scanpath is used along with the image to train our proposed classification model. An ordered set of fixated image region features as dictated by the generated scanpath is fed through an iterative sequential model (ISM) with an attention module to produce the radiologists' viewing pattern-related image features. These features are then employed along with visual features from the CXR image to perform the classification.

Our scanpath predictor is observed to be effective in generating human-like visual scanpaths when evaluated using the limited number of recorded scanpaths of radiologists available. Extensive experimentation shows that the use of generated scanpaths provides a performance improvement in automated multi-class multi-label disease classification considering both within-dataset and cross-dataset testing. This indicates that the scanpath guidance allows for better generalization capabilities of the learned classification model. We establish the importance of artificial visual scanpath usage by classifying $14$ pathological findings on large-scale data consisting of around $0.2$ million CXR images from $2$ different datasets, which to our best knowledge has not been attempted before.
The contributions and significance of this paper are as follows:

\begin{itemize}

\item To the best of our knowledge, this is the first work which shows that artificial visual scanpaths can improve the performance of multi-class multi-label disease classification in CXR images across datasets. 

\item To the best of our knowledge, our visual scanpath predictor is the first model to generate a radiologist's eye movement behavior on a CXR image for using it in automated screening. 


\item We propose a novel framework, which involves an iterative sequential model and an attention module, to incorporate a generated visual scanpath into the thoracic disease classification model for CXR images.

\end{itemize}

\section{Related Work}
\label{sec:related}

\begin{figure*}[htbp]
    \centering
    \includegraphics[width=0.95\textwidth]{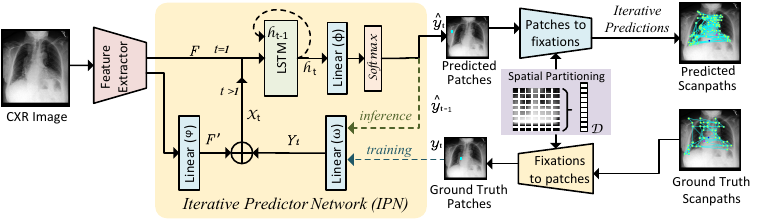}
    \caption{The proposed visual scanpath predictor network is trained on the recorded visual scanpaths from radiologists on CXR images. The spatial partitioning of images and forming the dictionary $\mathcal{D}$ convert the task of pixel-level prediction of fixation location to iterative patch classification. The iterative predictions of patches as fixation locations generate a visual scanpath on a given CXR image.}
    \label{fig: scan_pred}
\end{figure*}

\subsection{Automatic Disease Classification in Chest X-ray}
With the advent of deep learning and increased availability of chest X-Ray (CXR) datasets, designing automatic systems for CXR analysis for different diseases garnered immense attention in the research community~\cite{qin2018computer, ccalli2021deep, azizi2021big}. Among various works on CXR image analysis, a few attempts are aimed at disease detection and classification. Rajpurkar et al.~\cite{rajpurkar2017chexnet} developed one of the few initial deep models, ChexNet, to detect pneumonia on CXR.
Chen et al.~\cite{chen2022tae} introduced a generalized lung segmentation model using a tilewise autoencoder on CXR. Yan et al.~\cite{yan2018weakly}  employed a weakly supervised approach for different thoracic disease classifications. Wang et al.~\cite{wang2019thorax} proposed Thorax-Net, a Grad-CAM based attention-driven framework assisted by ResNet based classification network to diagnose different thorax diseases using CXR.
Ouyang et al.~\cite{ouyang2020learning} proposed a weakly-supervised approach based on a hierarchical attention framework for abnormality localization and classification on CXR. Bozorgtaba et al.~\cite{bozorgtabar2020salad} developed an end-to-end self-supervised model, Salad, that learns to detect anomalies on CXR. Wang et al.~\cite{wang2021automatically} designed a feature pyramid network (FPN) based model on top of ResNet to discriminate COVID-19 from common pneumonia and a residual attention network to localize the same. Paul et al.~\cite{paul2021discriminative} proposed an image triplets based few-shot learning framework for CXR screening. Wang et al.~\cite{wang2021triple} proposed a combination of three different attention modules built on top of a pretrained DenseNet-121 base to capture the features from CXR of different diseases. Kim et al.~\cite{kim2021xprotonet} developed a prototype matching-based network, that learns associated patterns of different diseases from CXR images. Agu et al.~\cite{agu2021anaxnet} developed graph convolutional networks based AnaXNnet that localizes anatomical regions of interest related to diseases. Zhou et al.~\cite{9316239} proposed two distinct abnormal attention models and a dual-weighting graph convolution technique to improve the accuracy of identifying multiple thoracic diseases on CXR. Mahapatra et al.~\cite{9361645} introduced a novel sample selection method for lung disease classification using interpretability saliency maps and self-supervised learning to identify informative samples without ground truth labels.
Recently, Liu et al.~\cite{liu2022acpl} proposed a semi-supervised medical image classification technique that exploits a pseudo-labeling mechanism. Feng et al.~\cite{feng2022contrastive} exploited contrastive learning for pneumonia and COVID-19 diagnosis. Mao et al.~\cite{9718305} introduced ImageGCN, a graph convolutional network framework designed to enhance image representations by considering image-level relations on chest X-ray images for disease identification.
Kamal et al.~\cite{kamal2022anatomy} proposed a CNN-based anatomy-aware network to classify different thoracic diseases. Han et al.~\cite{9930800} recently introduced a Radiomics-Guided Transformer (RGT) merging global image data with specific radiomic features to enhance cardiopulmonary pathology localization and classification in chest X-rays.

\subsection{Scanpath and Visual Attention}
\par Eye movement data is utilized in many computer vision tasks like object detection~\cite{yun2013studying, papadopoulos2014training}, image segmentation~\cite{mishra2009active, ramanathan2010eye}, action recognition in images~\cite{ge2015action} and videos~\cite{mathe2014actions}.
 
\par Recently, eye movement data has been utilized in the medical imaging field to improve the classification and localization of the disease. Ahmed et al.~\cite{ahmed2016fetal} proposed a fetal ultrasound image classification model that is trained on SURF descriptors extracted around eye fixation locations. In recent years, radiologists' eye movement data has been incorporated into the training of deep learning techniques to enhance disease classification.  
Karargyris et al.~\cite{karargyris2021creation} developed a dataset of CXR images that contain eye-tracking data from radiologists. They also demonstrated that the incorporation of eye gaze temporal information with CXR images in the disease classification model leads to improvement of the prediction task. Bhattacharya et al.~\cite{bhattacharya2022gazeradar} proposed GazeRadar which fused radiomic signatures with the representation of radiologists' visual search patterns for disease localization and binary classification.


\section{The Proposed Deep Models}
\label{sec:method}


\subsection{Scanpath Prediction Model}
\label{subsec:scan_pred}

As discussed earlier,  expert radiologists visually inspect  CXR images in a systematic way \cite{kok2016systematic}, which can be learnt from their recorded visual scanpaths on CXR images. As a recorded visual scanpath on a CXR image is not always available, we propose an iterative predictor network to predict scanpaths in CXR images. In this section, we discuss the architecture of the scanpath prediction network. As shown in Fig.~\ref{fig: scan_pred}, our model dynamically predicts a sequence of fixation locations for a given CXR image, which is then used in our proposed dynamic network for multi-class multi-label disease classification in the next section.

\par \textbf{Inputs and image partitioning:} As visual scanpath patterns of radiologists depend on the content of CXR images, we employ a CNN-based feature extractor to get the semantic feature $F$ that represents the visual content of an image. We resize and pad the CXR images to convert them into fixed-size square images. Furthermore, we spatially partition the square images into several small non-overlapping square patches and uniquely tag each patch with its spatial location in the image. A dictionary $\mathcal{D}$ is formed with all the patches as entries in it, each of which is associated with a unique key $k$. A token for the end of the sequence is also added to the dictionary $\mathcal{D}$. During training, fixation locations in a visual scanpath $(y = \left \{ y_{1},y_{2}, ..., y_{T} \right \})$ are binned into one of the patches transforming the task of scanpath prediction from pixel-level prediction to repeated classification into one of the patches. Hence, during inference, the predicted scanpath is a sequence of the patches in the input CXR image.

\begin{figure*}[!ht]
    \centering
    \includegraphics[width=0.95\textwidth]{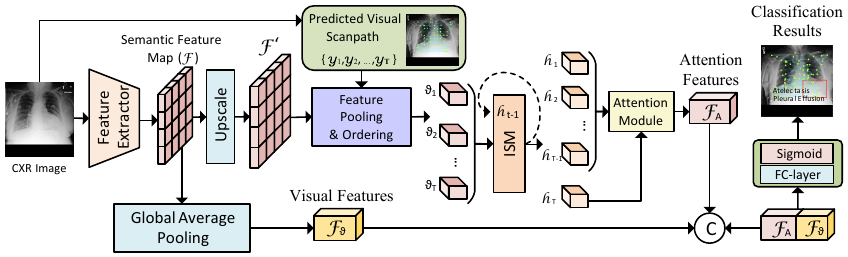}
    \caption{Visual scanpath-based dynamic network for multi-class multi-label disease classification. The network comprises of a CNN as a feature extractor and an iterative sequential model (ISM). In one branch of the network semantic feature map is passed through the global average pooling layer to produce visual features. The other branch employs ISM with an attention module on the sequence of visual features corresponding to a visual scanpath to give a viewing-pattern feature. The visual feature and viewing pattern feature are concatenated and fed to the classifier.}
    \label{fig: xray_class}
\end{figure*}

\par \textbf{Iterative predictor network (IPN) and output:} We design an iterative predictor network (IPN) to predict the current fixation location based on the previous fixation location and visual features of the input CXR image. The CXR image features represent the context during the above fixation predictions, except at the first fixation location prediction which is only based on the visual features of the input image.
The recurrent neural networks have been applied effectively in human behavior prediction, especially in processes involving sequential events~\cite{martinez2017human}. Thus, we use LSTM~\cite{hochreiter1997long} in our iterative predictor network (IPN). As shown in Fig.~\ref{fig: scan_pred}, for a given CXR image, the extracted visual feature $F$ alone is fed into the LSTM layer at the first time instance, and thereafter, $X_{t}$ is fed as the inputs at subsequent time instances to encode the spatio-temporal characteristics of visual scanpath. Here, $X_{t}$ is the concatenation of a lower-dimensional representation $F^{'}$ of the visual feature $F$ and an embedding $Y_{t}$ of the current ground truth fixation location $y_{t}$ (during training) or the previous fixation location $\hat{y}_{t-1}$ (during inference). A linear layer $\varphi$ is used to obtain the lower dimensional representation $F^{'}$ of visual features $F$. Similarly, another linear layer $\omega$ is employed to get the embedding $Y_{t}$ of a fixation location $y_{t}$ or $\hat{y}_{t-1}$. Given inputs $F$, $X_{t}$ and current hidden state $h_{t}$ of the LSTM, the probability distribution over all possible patches in the input image that depicts the probabilities of being the current fixation location is calculated as:
\begin{equation}
    p_{t} = softmax(\phi(h_{t};\theta_{\phi}))
\end{equation}
where $\phi$ is a linear layer with learnable parameters $\theta_{\phi}$. The patch with the highest probability in the probability distribution $p_{t}$ is selected as the predicted fixation location $y_{t}$ at time instance $t$.
\begin{equation}
    \hat{y}_{t} = \mathcal{D}[k^*],\   k^*=\textrm{argmax}_{k}\ p_{t}(k)
\end{equation}
where with the reference of the selected key $k^*$, the value, that is, the predicted fixation location $\hat{y}_{t}$ is fetched out from the dictionary $\mathcal{D}$.

\par \textbf{Loss function:} With the recursive prediction of patch/ fixation location till the detection of the end of the sequence token, the visual scanpath $(\hat{y} = \left\{ \hat{y}_{1},\hat{y}_{2}, ..., \hat{y}_{T} \right \})$ is predicted for the given CXR image. The summation of negative log-likelihood loss over all the time instances in a visual scanpath is used for the training of the proposed network:  
\begin{equation}
    L\left ( \theta  \right )= -\sum_{t=1}^{T}\log p_{t}\left ( k_{y_{t}} \right ) - \log p_{T+1}\left ( k_{e} \right )
\end{equation}
where $T$ is the length of ground truth scanpath $y$ and $k_{y_{t}}$ is the key in $\mathcal{D}$ for the ground truth fixation location at time instance $t$. $k_{e}$ is the key in $\mathcal{D}$ for the token of end of the sequence.

\subsection{Visual scanpath-based dynamic network for multi-class multi-label disease classification}
\label{subsec:scan_fusion}

 Studies~\cite{kok2016systematic} suggest that a systematic viewing pattern exists in the visual scanpaths of radiologists recorded while reviewing CXR images. This is because they are trained to examine clinically significant regions within CXR images and conduct diagnostic assessments. Therefore, the use of viewing patterns of radiologists along with CXR image features can aid disease classifiers in disease diagnosis on CXR images. In the previous section, we presented a  visual scanpath prediction model that can be trained on the recorded eye movement data of radiologists. We use the trained scanpath prediction model to generate visual scanpaths on arbitrary CXR images, which are not associated with recorded scanpaths. Therefore, we get CXR images with associated artificial visual scanpaths to be used in the training of our proposed dynamic network for multi-class multi-label disease classification. This allows the incorporation of radiologists' eye movement characteristics during the automated screening of CXR images for disease classification using our dynamic network.

\par As shown in Fig.~\ref{fig: xray_class}, we adopt a CNN network as a backbone to extract a semantic feature map $\mathcal{F}$ from a given CXR image. Then the network is bifurcated into two branches. In the first branch, the semantic feature map is passed through a global average pooling layer to produce visual features $\mathcal{F}_{v}$. In the second branch, to simulate the dynamic process of reviewing the CXR images by radiologists (reflected in their visual scanpaths), we employed an iterative sequential model (ISM) with an attention module~\cite{vaswani2017attention} which gives the dynamic viewing pattern features.  
  
\par Here, we describe the working of the iterative sequential model (ISM) with attention in detail. Let us denote a visual scanpath of a radiologist on a CXR image by $y = \left \{ y_{1},y_{2}, ..., y_{T} \right \}$ where $y_{t}, t=1,2,..., T$ represents the fixation locations in the image. We consider fixation locations as patches in CXR images representing square-shaped spatial regions. For a given image and its visual scanpath, first, we upscale the feature map $\mathcal{F}$ to $\mathcal{F}^{'}$. Then from the upscaled feature map $\mathcal{F}^{'}$ we pool out visual features $\vartheta_{t}$ corresponding to fixation locations/patches $y_{t}$ in the scanpath $y$ and arrange them in a sequence by maintaining the order of the corresponding fixation locations in the scanpath. We represent the sequence of visual features for the visual scanpath $y$ as $\vartheta = \left \{ \vartheta_{1}, \vartheta_{2}, ..., \vartheta_{T} \right \}$, which is fed into ISM that contains an LSTM layer. The expressions for the process at each time step are:
\begin{equation}
 h_{t} = f_{ISM}\left ( h_{t-1},\vartheta_{t} \right ) \; \; \; \;  0<t \leq T  
\end{equation}
where $T$ is the length of the sequence of visual features $\vartheta$. We apply an attention module to the hidden state $h_{t}$ of the LSTM layer (acting also as its output). In the module, the last hidden state $h_{T}$ attends to the hidden states at all the previous time instances and gives the viewing pattern features as:
\begin{equation}
    \mathcal{F}_{A} = f_{attn}(h_{T},\{h_{1},..., h_{T-1}\}, \{h_{1},..., h_{T-1}\})
\end{equation}
where $f_{attn}$ is an attention function~\cite{vaswani2017attention}. Finally, the visual features $\mathcal{F}_{\vartheta}$ and viewing pattern features $\mathcal{F}_{A}$ at the output of the two branches of the proposed network are concatenated and passed through a fully connected layer acting as a classifier to compute the score $S(d)$ for each disease category $d$. The computed score $S(d)$ for each disease category is then passed through the sigmoid function to give the estimated probability $\hat{p}_d$ of the disease category $d$.

\par \textbf{Loss function:} There are two challenges in handling labels for disease classification in CXR images~\cite{rajpurkar2017chexnet, johnson2019mimic}. The first one is the class imbalance issue and the second one is the presence of uncertain labels. There are labels where radiologists have confidently stated whether the disease is present or not. However, there are many cases where radiologists are not confident about the existence of a disease (a label). Such uncertain labels are usually mentioned in CXR datasets. 
\textcolor{black}{As uncertain labels may push the neural network training in the wrong direction, we mask those labels during training by excluding them from loss calculations and model weight updation.} 

To handle the class-imbalance problem, we use a reverse weighing strategy during the loss computation. For example, if class $d$ is present in $n_{pos}$ number of images and absent in $n_{neg}$ number of images, we assign weights $w_{pos}=n_{neg}/(n_{pos}+n_{neg})$ when class $d$ is encountered. Generally, in thoracic disease datasets $n_{neg}>>n_{pos}$. Therefore, we give lesser weight ($1-w_{pos}$) in the loss function during training if that disease is absent in the image at hand and comparably more weight ($w_{pos}$) if the disease is present. We modify the cross-entropy loss and formulate `uncertainty-removed weighted cross-entropy loss' to train the networks. It can be mathematically expressed as
\begin{equation}
    \mathcal{L}(p_d, \hat{p}_d) = \sum\{p_d\neq u\}w_d[p_d\log \hat{p}_d+(1-p_d)\log (1-\hat{p}_d)].
\end{equation}
where, $p_d$ is the actual label of class $d$. $\hat{p}_d$ is the predicted probability of class $d$. $w_d$ is the weight for class $d$. $u$ represents the presence of uncertainty in that label. The weight for class $d$ is expressed as
\begin{equation}
    w_d = p_d w_{pos}+(1-p_d)(1-w_{pos}).
\end{equation}





\section{Results and Discussion}
\label{sec:results}

\subsection{Datasets}
\label{subsec:data}

There are two popular publicly available datasets, MIMIC~\cite{johnson2019mimic} and CheXpert~\cite{irvin2019chexpert} for thoracic disease classification containing CXR images in the order of one-tenth of a million. Both the datasets are labeled for $14$ different thoracic diseases. REFLACX dataset~\cite{lanfredi2022reflacx}, which contains visual scanpath data of radiologists recorded while they were performing diagnosis, is created for a small subset of images from the MIMIC dataset. There is a total of 2616 images where visual scanpath data is available. In a few cases, multiple radiologists' visual scanpath data are available for a single CXR image, where we randomly select one. 
\textcolor{black}{In the first stage, to train our scanpath prediction model, we consider the train, validation and test splits containing 2084, 10, and 522 images, respectively, as provided in REFLACX dataset~\cite{lanfredi2022reflacx}, which reflects the corresponding splits in MIMIC dataset~\cite{johnson2019mimic}.
All the eye-tracking data in the REFLACX dataset is collected on CXR images tagged with ``ViewPosition" metadata values. These values are either Posterior Anterior (PA) or Anterior Posterior (AP), which indicates whether the patient's chest was facing the X-ray detector or the X-ray machine.}
Therefore, we only consider CXR images with frontal views (PA/ AP) from MIMIC and CheXpert datasets for classification. Among the $243,324$ frontal images from the MIMIC dataset, we use $237,962$ ($MIMIC\_Train$) for training our multi-class multi-label classification model. $1,959$ ($MIMIC\_Valid$) images are used for validation and $3,403$ ($MIMIC\_Test$) for testing. The train, validation and test split is already provided in the MIMIC dataset. We also use $191,010$ frontal images from the CheXpert training dataset ($CheXpert\_Train$) in experiments to test our model.

\subsection{Experimental Settings} \label{ExperimentalSettings}
\textcolor{black}{In both our proposed models, we pad zeros to the shorter side of an input CXR image to form a square image without distorting its contents. Then, we resize the image to $256\times256$ using bilinear interpolation.}

\par In the proposed visual scanpath predictor, we employ Imagenet pretrained ResNet-152 as the feature extractor by removing its last fully connected linear layer to extract a semantic feature $F$. As shown in Fig.~\ref{fig: scan_pred}, the linear layer $\varphi$ converts feature $F$ to $F'$ into a $256$ dimensional vector. The Iterative predictor network (IPN) used a single-layer LSTM with $512$ hidden units. The linear layer $\omega$ embeds patches into $256$ dimensional feature vector, whereas linear layer $\phi$ converts hidden states into $257$ dimensional vector representing scores corresponding to each entry in the dictionary $\mathcal{D}$. The model is trained for $200$ epochs with Adam optimizer, and the learning rate is set to $10^{-5}$.

Most of the automatic X-ray analysis  frameworks adopt DenseNet~\cite{huang2017densely} or ResNet~\cite{he2016deep}  architecture as the baseline /backbone because of their promising results~\cite{ouyang2020learning, ccalli2021deep, guan2020multi, kim2021xprotonet, li2018thoracic, liu2022acpl, paul2021generalized}.  Therefore, we follow the same and consider ResNet-18 and ResNet-50 in ResNet architecture, and DenseNet-121 and DenseNet-201 in DenseNet architecture as the backbone of our proposed scanpath-based dynamic network for multi-label disease classification as explained in Section~\ref{subsec:scan_fusion}. We use the same experimental settings in all the experiments for fair comparative analysis. For simplicity, we do not use any data augmentation method. We train the network for $25$ epochs, and the learning rate is decreased by a factor of $0.2$ after every $6$ epoch. All the networks are trained by using Adam optimizer with an initial learning rate $10^{-4}$ and batch size $16$. The models which give the best validation AUROC are chosen for testing purposes and performance evaluation. We train the model on $MIMIC\_Train$ dataset and $MIMIC\_Valid$ is used for validation purposes. $MIMIC\_Test$ and the huge $CheXpert\_Train$ datasets are used for testing, with the latter signifying cross-dataset testing.

\subsection{Performance of Scanpath Prediction}

\par We use two established measures, MultiMatch~\cite{dewhurst2012depends} and ScanMatch~\cite{cristino2010scanmatch}, to evaluate the scanpath prediction model. These measures consider spatial and ordinal similarities between fixation locations of two scanpaths of any given length by normalizing the measures according to sequence lengths.

{\color{black}MultiMatch measure computes the similarity between two scanpaths in terms of five sub-metrics related to their vectors, lengths, directions, positions, and durations~\cite{dewhurst2012depends}. The sequences of fixations in two scanpaths being compared are aligned using dynamic programming based on scanpath shapes and then the sub-metrics are computed separately. The vector sub-metric quantifies the normalized difference between aligned pairs of saccade vectors to measure the resemblance of scanpath shapes. The length sub-metric measures the normalized difference in lengths between aligned saccades, indicating similarity in saccadic amplitudes. The direction sub-metric evaluates the normalized angular disparity of saccade vectors, again measuring shape similarity. The position sub-metric assesses the normalized positional difference between aligned fixations, reflecting scanpath similarity in terms of spatial locations. As we are not predicting the fixation duration, the duration sub-metric does not apply to this paper. Note that it was pointed out in \cite{sun2019visual} that as only aligned parts contribute to the similarity measurement in MultiMatch, seemingly high similarity values may be achieved in cases where the similarity in the scanpaths is not obvious visually.

ScanMatch, a method for comparing scanpaths based on regions of interest (ROI), employs the Needleman–Wunsch algorithm~\cite{Needleman1970} for alignment and comparison~\cite{cristino2010scanmatch}. Initially, in ScanMatch, scanpaths undergo encoding into character strings by allocating fixations into evenly distributed bins across the image space, with each bin represented by a unique pair of letters. The comparison of two scanpath strings involves maximizing the similarity score computed from a substitution matrix, which determines the score for all possible letter pair substitutions based on the corresponding spatial bin location in the image space along with a gap penalty. The ScanMatch measure is calculated without considering fixation duration, as it is not applicable in our case.}


\begin{table}[]
\centering
\caption{\color{black} Performance comparison of the proposed scanpath prediction model with Radiologist, GenLSTM, Random and CLE on the test set of REFLACX dataset.}
\label{tab:T1}
\scalebox{0.9}{
\begin{tabular}{l@{\hskip .12cm}l@{\hskip .16cm}c@{\hskip .12cm}c@{\hskip .12cm}c@{\hskip .12cm}c@{\hskip .12cm}c}
\cmidrule(){1-7}
\multirow{2}{*}{\begin{tabular}[c]{@{}l@{}}Image\\ Set\end{tabular}} & \multicolumn{1}{l}{\multirow{2}{*}{Method}} & \multicolumn{4}{c}{MultiMatch Scores $\uparrow$}  & ScanMatch $\uparrow$      \\ \cmidrule(){3-6} 
                                                                     & \multicolumn{1}{c}{}                        & Vector & Direction & Length & Position & (w/o Duration) \\ \cmidrule(lr){1-7}
\multirow{4}{*}{Set-1}                                               & Radiologist                                       & 0.9521 & 0.7358    & 0.9490 & 0.8243   & 0.3136         \\ \cmidrule(r){2-2} \cmidrule(r){3-6} \cmidrule(r){7-7}                                              & Random                                     & 0.8703 & 0.7161    & 0.8146 & 0.7312   & 0.2673         \\ 
                                                                     & CLE~\cite{Boccignone2004}                                         & \textbf{0.9654} & 0.6345    & 0.9364 & 0.6657   & 0.1288         \\
                                                                     & GenLSTM~\cite{verma2024generative}                                     & 0.9546 & 0.7339    & 0.9523 & 0.8028   & 0.3102         \\ 
                                                                      \cmidrule(r){2-2} \cmidrule(r){3-6} \cmidrule(r){7-7} 
                                                                     & Our                                         & 0.9538 & \textbf{0.7350}    & \textbf{0.9552} & \textbf{0.8038}   & \textbf{0.3127}         \\ \cmidrule(lr){1-7}
\multirow{3}{*}{Set-2}
                                                                     & Random                                     & 0.8702 & 0.7154    & 0.8115 & 0.7298   & 0.2853         \\                                               & CLE~\cite{Boccignone2004}                                         & \textbf{0.9676} & 0.6286    & 0.9406 & 0.6459   & 0.1186         \\
                                                                     & GenLSTM~\cite{verma2024generative}                                     & 0.9564 & 0.7338    & 0.9528 & 0.8129   & 0.3408         \\ \cmidrule(r){2-2} \cmidrule(r){3-6} \cmidrule(r){7-7} 
                                                                     & Our                                         & 0.9539 & \textbf{0.7376}    & \textbf{0.9537} & \textbf{0.8145}   & \textbf{0.3513}         \\ \cmidrule(){1-7}
\end{tabular}}
\end{table}

\begin{figure*}[htbp]
    \centering
    \includegraphics[width=0.825\textwidth]{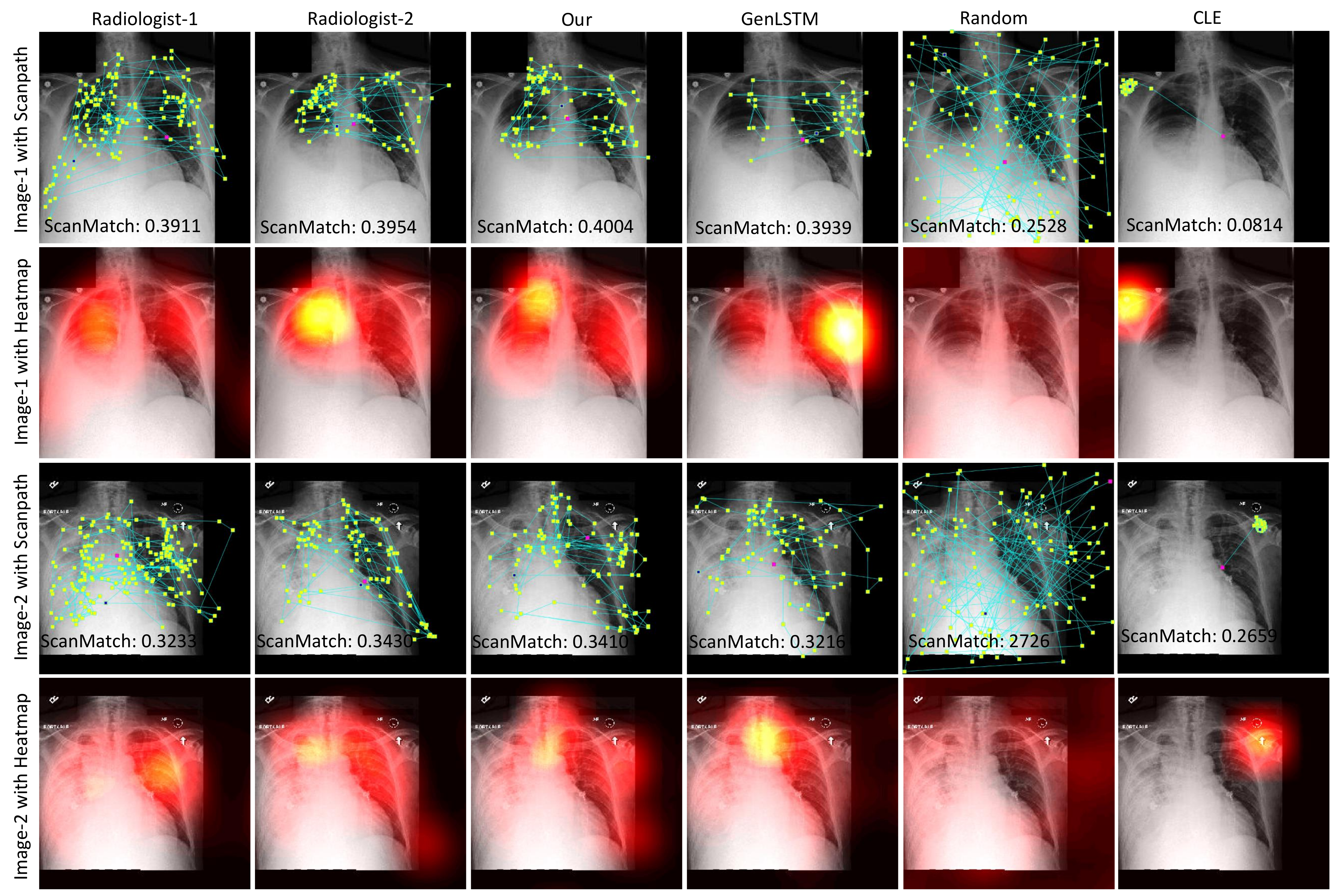}
    \caption{\color{black} Performance of proposed visual scanpath prediction model compared to two radiologists (Radiologist-1 and Radiologist-2), GenLSTM, Random and CLE on two images (Image-1 and Image-2). The first and third rows plot the scanpaths with corresponding ScanMatch values on Image-1 and Image-2. The second and fourth rows plot the heatmaps corresponding to the scanpaths plotted in the first and third rows, respectively.}
    \label{fig:scan_plot}
\end{figure*}

\par To evaluate the generation of visual scanpaths, we compare the scanpath prediction performance of our proposed model with that of Radiologist scanpaths, randomly generated scanpaths and two other methods. Based on the available number of scanpaths on each CXR image, we evaluate the performance on two sets of the test images of the REFLACX dataset. Let us represent the first set as `Set-1' which is the subset of the test set of the REFLACX dataset, where scanpaths are available for multiple radiologists on each CXR image. In this set, the performance measures for a single scanpath generated by a given model on a CXR image are calculated by averaging the related measures computed against all the radiologists' scanpaths on that image. The second set `Set-2' contains all the test CXR images of the REFLACX dataset. However, on an image having scanpaths of multiple radiologists, only one is randomly chosen for the performance evaluation. Therefore, in this set, the performance measures for a single scanpath generated by a given model are computed against the corresponding solitary radiologist's scanpath.

\par The Radiologist performance indicates the similarity among scanpaths of multiple radiologists on the same CXR image. Hence, it can be calculated on the `Set-1' images only and may be considered as a benchmark for performance comparison. To calculate the value of the Radiologist performance measure, we compute the measures for each radiologist's scanpath on the CXR image against all other radiologists' scanpaths on that image. Then, the calculated values for all the radiologists' scanpaths on that image are averaged to obtain the Radiologist performance on that image. The `Random' model is considered to compare against results obtained without any design. We generate one random scanpath on each CXR image in the REFLACX test set by randomly selecting $N$ number of locations (equally likely) in the given image as fixation locations. We select $N$ equal to the average scanpath length in the REFLACX training set. To the best of our knowledge, no other scanpath prediction model for CXR images is available in the literature. Our model is trained to generate a scanpath directly from an input CXR image. 
As performed by the CLE method of \cite{Boccignone2004}, the estimate of human fixations in the form of a saliency map could also be used to compute a visual scanpath. Therefore, we consider the performance comparison of our scanpath prediction model with CLE as well.
\textcolor{black}{We also include a deep learning-based scanpath prediction method, GenLSTM~\cite{verma2024generative}, in our comparisons.}

\begin{table*}[htbp]
 \centering
 \caption{The classification performance of our proposed approach in terms of AUROC and AUPRC using ResNet and DenseNet feature extractor architectures with and without scanpath guidance on $MIMIC\_Test$ and $CheXpert\_Train$ datasets.}
\label{tab:T2}
 \scalebox{0.88}{
\begin{tabular}{l@{\hskip 0.12cm}|c@{\hskip 0.12cm}c@{\hskip 0.12cm}c@{\hskip 0.12cm}c@{\hskip 0.12cm}c@{\hskip 0.12cm}c@{\hskip 0.12cm}|c@{\hskip 0.12cm}c@{\hskip 0.12cm}c@{\hskip 0.12cm}c@{\hskip 0.12cm}c@{\hskip 0.12cm}l}
\cmidrule(lr){1-13}
Dataset       & \multicolumn{6}{c|}{$MIMIC\_Test$ (Within Dataset) }                                                                                                                                                                                                                                                                                                                                                                                                                                              & \multicolumn{6}{c}{$CheXpert\_Train$ (Cross Dataset)}                                                                                                                                                                                                                                                                                                                                                                                                                                                 \\ \cmidrule(lr){1-1}\cmidrule(lr){2-7}\cmidrule(lr){8-13}
Measures      & \multicolumn{3}{c|}{AUROC}                                                                                                                                                                                                            & \multicolumn{3}{c|}{AUPRC}                                                                                                                                                                                                            & \multicolumn{3}{c|}{AUROC}                                                                                                                                                                                                            & \multicolumn{3}{c}{AUPRC}                                                                                                                                                                                                            \\ \cmidrule(lr){1-1}\cmidrule(lr){2-4}\cmidrule(lr){5-7}\cmidrule(lr){8-10}\cmidrule(lr){11-13}
\begin{tabular}[l]{@{}l@{}}Base \\  Network \end{tabular} & \multicolumn{1}{c|}{\begin{tabular}[c]{@{}c@{}c@{}}w/o\\ scanpath \\ (A) \end{tabular}} & \multicolumn{1}{c|}{\begin{tabular}[c]{@{}c@{}c@{}}with\\ scanpath \\ (B)\end{tabular}} & \multicolumn{1}{c|}{\begin{tabular}[c]{@{}c@{}c@{}}\% imprnt\\ $\frac{\left | A-B \right |}{A}$ \tiny x100 \end{tabular}} & \multicolumn{1}{c|}{\begin{tabular}[c]{@{}c@{}c@{}}w/o\\ scanpath \\ (A)\end{tabular}} & \multicolumn{1}{c|}{\begin{tabular}[c]{@{}c@{}c@{}}with\\ scanpath \\ (B)\end{tabular}} & \multicolumn{1}{c|}{\begin{tabular}[c]{@{}c@{}c@{}}\% imprnt\\ $\frac{\left | A-B \right |}{A}$ \tiny x100\end{tabular}} & \multicolumn{1}{c|}{\begin{tabular}[c]{@{}c@{}c@{}}w/o\\ scanpath \\ (A)\end{tabular}} & \multicolumn{1}{c|}{\begin{tabular}[c]{@{}c@{}c@{}}with\\ scanpath \\ (B)\end{tabular}} & \multicolumn{1}{c|}{\begin{tabular}[c]{@{}c@{}c@{}}\% imprnt\\ $\frac{\left | A-B \right |}{A}$ \tiny x100\end{tabular}} & \multicolumn{1}{c|}{\begin{tabular}[c]{@{}c@{}c@{}}w/o\\ scanpath \\ (A)\end{tabular}} & \multicolumn{1}{c|}{\begin{tabular}[c]{@{}c@{}c@{}}with\\ scanpath \\ (B)\end{tabular}} & \multicolumn{1}{c}{\begin{tabular}[c]{@{}c@{}c@{}}\% imprnt\\ $\frac{\left | A-B \right |}{A}$ \tiny x100\end{tabular}} \\  \cmidrule(lr){1-1} \cmidrule(lr){2-2} \cmidrule(lr){3-3} \cmidrule(lr){4-4} \cmidrule(lr){5-5} \cmidrule(lr){6-6} \cmidrule(lr){7-7} \cmidrule(lr){8-8} \cmidrule(lr){9-9} \cmidrule(lr){10-10} \cmidrule(lr){11-11} \cmidrule(lr){12-12} \cmidrule(lr){13-13}
ResNet-18     & 0.7991                                                                      & 0.8003                                                                       & \multicolumn{1}{l|}{\textbf{0.1492}}                                     & 0.4539                                                                      & 0.4573                                                                       & \textbf{0.7367}                                                          & 0.7195                                                                      & 0.7444                                                                       & \multicolumn{1}{l|}{\textbf{3.4593}}                                     & 0.3605                                                                      & 0.4022                                                                       & \textbf{11.5720}                                                        \\
ResNet-50     & 0.7928                                                                      & 0.8045                                                                       & \multicolumn{1}{l|}{\textbf{1.4650}}                                     & 0.4598                                                                      & 0.4717                                                                       & \textbf{2.5798}                                                          & 0.6637                                                                      & 0.6898                                                                       & \multicolumn{1}{l|}{\textbf{3.9402}}                                     & 0.2998                                                                      & 0.3283                                                                       & \textbf{9.5078}                                                         \\
DenseNet-121      & 0.8028                                                                      & 0.8012                                                                       & \multicolumn{1}{l|}{-0.2012}                                             & 0.4615                                                                      & 0.4626                                                                       & \textbf{0.2301}                                                          & 0.6356                                                                      & 0.6687                                                                       & \multicolumn{1}{l|}{\textbf{5.2050}}                                     & 0.2615                                                                      & 0.3126                                                                       & \textbf{19.5519}                                                        \\
DenseNet-201      & 0.8061                                                                      & 0.8099                                                                       & \multicolumn{1}{l|}{\textbf{0.4664}}                                     & 0.4671                                                                      & 0.4767                                                                       & \textbf{2.0474}                                                          & 0.6998                                                                      & 0.7039                                                                       & \multicolumn{1}{l|}{\textbf{0.5783}}                                     & 0.3210                                                                      & 0.3581                                                                       & \textbf{11.5806} \\ \cmidrule(lr){1-13}                                                      
\end{tabular}}
\vspace{-4pt}
\end{table*}

\par {\color{black}The performance comparison based on MultiMatch and ScanMatch measures is shown in Table~\ref{tab:T1}. For every method in the table, we generate a single scanpath on each CXR image in the test set of the REFLACX dataset. The performance measures are computed for the single generated scanpath by `Our', `Random', `CLE' and `GenLSTM' methods in each CXR image of `Set-1' and `Set-2'. The values shown in Table~\ref{tab:T1} are the average of the measure values for all the CXR images in `Set-1' and `Set-2'. From `Set-1' in Table~\ref{tab:T1}, we observe that the measure values for our proposed method on both the image sets are very close to the Radiologist performance. Our method also outperforms CLE and GenLSTM in most cases of both the `Sets' and is comfortably better than Random in all the cases.}


\par \textcolor{black}{The qualitative results are shown in Fig.~\ref{fig:scan_plot} with the help \textcolor{black}{of scanpaths of two different radiologists and that of the four computational methods from Table~\ref{tab:T1} on two CXR images. The ScanMatch measure values corresponding to the plotted scanpaths are also shown. The heatmaps corresponding to the scanpaths of the different methods on the two images are shown as well. From the figure, we observe that the recorded scanpaths generated by our approach are much closer to that of the two radiologists in comparison to the scanpaths of `GenLSTM', `Random' and CLE. Our method is trained to learn the strategic viewing patterns of radiologists in general, and hence, we observe that the scanpaths from our method vary from that of the two radiologists as much as they do from each other while successfully classifying diseases~\cite{lanfredi2022reflacx}.}}

\begin{table*}[htbp]
\centering
\caption{The classwise classification performance of our proposed approach in terms of AUROC and AUPRC using ResNet-50 as the backbone architecture in both the within-dataset and cross-dataset cases.}
\label{tab:T3}
\scalebox{0.85}{
\begin{tabular}{l@{\hskip 0.15cm}|c@{\hskip 0.15cm}c@{\hskip 0.15cm}c@{\hskip 0.15cm}c@{\hskip 0.15cm}c@{\hskip 0.15cm}c@{\hskip 0.15cm}|c@{\hskip 0.15cm}c@{\hskip 0.15cm}c@{\hskip 0.15cm}c@{\hskip 0.15cm}c@{\hskip 0.15cm}c}
\cmidrule(lr){1-13}
Dataset                                                                  & \multicolumn{6}{c|}{$MIMIC\_Test$  (Within Dataset)}                                                                                                                                                                                                                                                                                                                                                                                                                                              & \multicolumn{6}{c}{$CheXpert\_Train$  (Cross Dataset)}                                                                                                                                                                                                                                                                                                                                                                                                                                                 \\ \cmidrule(lr){1-1}\cmidrule(lr){2-7}\cmidrule(lr){8-13}
Measures                                                                 & \multicolumn{3}{c|}{AUROC}                                                                                                                                                                                                            & \multicolumn{3}{c|}{AUPRC}                                                                                                                                                                                                            & \multicolumn{3}{c|}{AUROC}                                                                                                                                                                                                            & \multicolumn{3}{c}{AUPRC}                                                                                                                                                                                                            \\ \cmidrule(lr){1-1}\cmidrule(lr){2-4}\cmidrule(lr){5-7}\cmidrule(lr){8-10}\cmidrule(lr){11-13}
Diseases                                                                 & \multicolumn{1}{c|}{\begin{tabular}[c]{@{}c@{}c@{}}w/o\\ scanpath \\ (A)\end{tabular}} & \multicolumn{1}{c|}{\begin{tabular}[c]{@{}c@{}c@{}}with\\ scanpath \\ (B)\end{tabular}} & \multicolumn{1}{c|}{\begin{tabular}[c]{@{}c@{}c@{}}\% imprnt\\ $\frac{\left | A-B \right |}{A}$ \tiny x100 \end{tabular}} & \multicolumn{1}{c|}{\begin{tabular}[c]{@{}c@{}c@{}}w/o\\ scanpath \\ (A)\end{tabular}} & \multicolumn{1}{c|}{\begin{tabular}[c]{@{}c@{}c@{}}with\\ scanpath \\ (B)\end{tabular}} & \multicolumn{1}{c|}{\begin{tabular}[c]{@{}c@{}c@{}}\% imprnt\\ $\frac{\left | A-B \right |}{A}$ \tiny x100 \end{tabular}} & \multicolumn{1}{c|}{\begin{tabular}[c]{@{}c@{}c@{}}w/o\\ scanpath \\ (A)\end{tabular}} & \multicolumn{1}{c|}{\begin{tabular}[c]{@{}c@{}c@{}}with\\ scanpath \\ (B)\end{tabular}} & \multicolumn{1}{c|}{\begin{tabular}[c]{@{}c@{}c@{}}\% imprnt\\ $\frac{\left | A-B \right |}{A}$ \tiny x100 \end{tabular}} & \multicolumn{1}{c|}{\begin{tabular}[c]{@{}c@{}c@{}}w/o\\ scanpath \\ (A)\end{tabular}} & \multicolumn{1}{c|}{\begin{tabular}[c]{@{}c@{}c@{}}with\\ scanpath \\ (B)\end{tabular}} & \multicolumn{1}{c}{\begin{tabular}[c]{@{}c@{}c@{}}\% imprnt\\ $\frac{\left | A-B \right |}{A}$ \tiny x100 \end{tabular}} \\ \cmidrule(lr){1-1} \cmidrule(lr){2-2} \cmidrule(lr){3-3} \cmidrule(lr){4-4} \cmidrule(lr){5-5} \cmidrule(lr){6-6} \cmidrule(lr){7-7} \cmidrule(lr){8-8} \cmidrule(lr){9-9} \cmidrule(lr){10-10} \cmidrule(lr){11-11} \cmidrule(lr){12-12} \cmidrule(lr){13-13}
Atelectasis:                                                             & 0.7493                                                                      & 0.7578                                                                       & \multicolumn{1}{l|}{\textbf{1.1381}}                                     & 0.4289                                                                      & 0.4439                                                                       & \textbf{3.5088}                                                          & 0.6263                                                                      & 0.6082                                                                       & \multicolumn{1}{l|}{-2.8754}                                             & 0.2524                                                                      & 0.2357                                                                       & -6.5921                                                                 \\
Cardiomegaly:                                                            & 0.7634                                                                      & 0.7663                                                                       & \multicolumn{1}{l|}{\textbf{0.3833}}                                     & 0.4859                                                                      & 0.4900                                                                       & \textbf{0.8521}                                                          & 0.6130                                                                      & 0.6561                                                                       & \multicolumn{1}{l|}{\textbf{7.0223}}                                     & 0.2620                                                                      & 0.2939                                                                       & \textbf{12.1974}                                                        \\
Consolidation:                                                           & 0.7751                                                                      & 0.7882                                                                       & \multicolumn{1}{l|}{\textbf{1.6954}}                                     & 0.4319                                                                      & 0.4401                                                                       & \textbf{1.8958}                                                          & 0.6480                                                                      & 0.6793                                                                       & \multicolumn{1}{l|}{\textbf{4.8420}}                                     & 0.2402                                                                      & 0.2681                                                                       & \textbf{11.6405}                                                        \\
Edema:                                                                   & 0.7956                                                                      & 0.8038                                                                       & \multicolumn{1}{l|}{\textbf{1.0347}}                                     & 0.4909                                                                      & 0.4917                                                                       & \textbf{0.1615}                                                          & 0.7097                                                                      & 0.7368                                                                       & \multicolumn{1}{l|}{\textbf{3.8179}}                                     & 0.3536                                                                      & 0.3965                                                                       & \textbf{12.1466}                                                        \\
\begin{tabular}[c]{@{}l@{}}Enlarged   \\ Cardiomediastinum:\end{tabular} & 0.7947                                                                      & 0.8143                                                                       & \multicolumn{1}{l|}{\textbf{2.4616}}                                     & 0.4617                                                                      & 0.4738                                                                       & \textbf{2.6197}                                                          & 0.6448                                                                      & 0.6842                                                                       & \multicolumn{1}{l|}{\textbf{6.1156}}                                     & 0.2729                                                                      & 0.3257                                                                       & \textbf{19.3317}                                                        \\
Fracture:                                                                & 0.8124                                                                      & 0.8256                                                                       & \multicolumn{1}{l|}{\textbf{1.6251}}                                     & 0.4516                                                                      & 0.4609                                                                       & \textbf{2.0515}                                                          & 0.6624                                                                      & 0.6846                                                                       & \multicolumn{1}{l|}{\textbf{3.3538}}                                     & 0.2610                                                                      & 0.3069                                                                       & \textbf{17.5749}                                                        \\
Lung Lesion:                                                             & 0.8144                                                                      & 0.8229                                                                       & \multicolumn{1}{l|}{\textbf{1.0489}}                                     & 0.4328                                                                      & 0.4321                                                                       & -0.1613                                                                  & 0.6889                                                                      & 0.7010                                                                       & \multicolumn{1}{l|}{\textbf{1.7522}}                                     & 0.2543                                                                      & 0.2971                                                                       & \textbf{16.8334}                                                        \\
Lung Opacity:                                                            & 0.7963                                                                      & 0.8077                                                                       & \multicolumn{1}{l|}{\textbf{1.4280}}                                     & 0.4282                                                                      & 0.4338                                                                       & \textbf{1.3152}                                                          & 0.6092                                                                      & 0.6383                                                                       & \multicolumn{1}{l|}{\textbf{4.7798}}                                     & 0.2422                                                                      & 0.2786                                                                       & \textbf{15.0033}                                                        \\
No Finding:                                                              & 0.7880                                                                      & 0.7935                                                                       & \multicolumn{1}{l|}{\textbf{0.7077}}                                     & 0.4318                                                                      & 0.4334                                                                       & \textbf{0.3713}                                                          & 0.6223                                                                      & 0.6644                                                                       & \multicolumn{1}{l|}{\textbf{6.7652}}                                     & 0.2461                                                                      & 0.2795                                                                       & \textbf{13.5566}                                                        \\
Pleural Effusion:                                                        & 0.8077                                                                      & 0.8133                                                                       & \multicolumn{1}{l|}{\textbf{0.6880}}                                     & 0.5106                                                                      & 0.5175                                                                       & \textbf{1.3648}                                                          & 0.6747                                                                      & 0.7045                                                                       & \multicolumn{1}{l|}{\textbf{4.4113}}                                     & 0.3532                                                                      & 0.3895                                                                       & \textbf{10.2862}                                                        \\
Pleural Other:                                                           & 0.7992                                                                      & 0.8151                                                                       & \multicolumn{1}{l|}{\textbf{1.9828}}                                     & 0.4738                                                                      & 0.4962                                                                       & \textbf{4.7302}                                                          & 0.6729                                                                      & 0.7130                                                                       & \multicolumn{1}{l|}{\textbf{5.9603}}                                     & 0.3375                                                                      & 0.3788                                                                       & \textbf{12.2243}                                                        \\
Pneumonia:                                                               & 0.7934                                                                      & 0.8087                                                                       & \multicolumn{1}{l|}{\textbf{1.9165}}                                     & 0.4623                                                                      & 0.4820                                                                       & \textbf{4.2591}                                                          & 0.6903                                                                      & 0.7274                                                                       & \multicolumn{1}{l|}{\textbf{5.3838}}                                     & 0.3341                                                                      & 0.3751                                                                       & \textbf{12.2544}                                                        \\
Pneumothorax:                                                            & 0.7974                                                                      & 0.8159                                                                       & \multicolumn{1}{l|}{\textbf{2.3184}}                                     & 0.4426                                                                      & 0.4743                                                                       & \textbf{7.1681}                                                          & 0.6895                                                                      & 0.7274                                                                       & \multicolumn{1}{l|}{\textbf{5.4911}}                                     & 0.3200                                                                      & 0.3675                                                                       & \textbf{14.8382}                                                        \\
Support Devices:                                                         & 0.8129                                                                      & 0.8293                                                                       & \multicolumn{1}{l|}{\textbf{2.0220}}                                     & 0.5049                                                                      & 0.5341                                                                       & \textbf{5.7887}                                                          & 0.7395                                                                      & 0.7323                                                                       & \multicolumn{1}{l|}{-0.9788}                                             & 0.4677                                                                      & 0.4034                                                                       & -13.7554                                                                \\ \cmidrule(lr){1-1} \cmidrule(lr){2-2} \cmidrule(lr){3-3} \cmidrule(lr){4-4} \cmidrule(lr){5-5} \cmidrule(lr){6-6} \cmidrule(lr){7-7} \cmidrule(lr){8-8} \cmidrule(lr){9-9} \cmidrule(lr){10-10} \cmidrule(lr){11-11} \cmidrule(lr){12-12} \cmidrule(lr){13-13}
Average                                                                  & 0.7928                                                                      & 0.8045                                                                       & \multicolumn{1}{l|}{\textbf{1.4650}}                                     & 0.4598                                                                      & 0.4717                                                                       & \textbf{2.5798}                                                          & 0.6637                                                                      & 0.6898                                                                       & \multicolumn{1}{l|}{\textbf{3.9402}}                                     & 0.2998                                                                      & 0.3283                                                                       & \textbf{9.5078}                                                         \\ \cmidrule(lr){1-13}
\end{tabular}}
\end{table*}

\begin{table*}[htbp]
\centering
\caption{\color{black} The classification performance of our proposed scanpath based classification approach with the guidance of scanpaths generated from our and a few other scanpath prediction models in terms of AUROC and AUPRC.}
\label{tab:T4}
\scalebox{0.83}{

\begin{tabular}{l@{\hskip 0.15cm}|c@{\hskip 0.15cm}c@{\hskip 0.15cm}c@{\hskip 0.15cm}c@{\hskip 0.15cm}|c@{\hskip 0.15cm}c@{\hskip 0.15cm}c@{\hskip 0.15cm}c@{\hskip 0.15cm}|c@{\hskip 0.15cm}c@{\hskip 0.15cm}c@{\hskip 0.15cm}c@{\hskip 0.15cm}|c@{\hskip 0.15cm}c@{\hskip 0.15cm}c@{\hskip 0.15cm}c}
\cmidrule(r){1-17}
Dataset      & \multicolumn{8}{c}{$MIMIC\_Test$ (Within Dataset)}                                                                                                                                                                      & \multicolumn{8}{c}{$CheXpert\_Train$ (Cross Dataset)}                                                                                                                                                                   \\ \cmidrule(r){1-1} \cmidrule(r){2-9} \cmidrule(r){10-17}
Measures     & \multicolumn{4}{c|}{AUROC}                                                                                & \multicolumn{4}{c|}{AUPRC}                                                                                  & \multicolumn{4}{c|}{AUROC}                                                                                & \multicolumn{4}{c}{AUPRC}                                                                                  \\ \cmidrule(r){1-1} \cmidrule(r){2-5} \cmidrule(r){6-9} \cmidrule(r){10-13} \cmidrule(r){14-17}
Base         & \multicolumn{1}{c|}{w/o}                        & \multicolumn{3}{c|}{with scanpath}                                           & \multicolumn{1}{c|}{w/o}      & \multicolumn{3}{c|}{with scanpath}                                           & \multicolumn{1}{c|}{w/o}                        & \multicolumn{3}{c|}{with scanpath}                                           & \multicolumn{1}{c|}{w/o}      & \multicolumn{3}{c}{with scanpath}                                           \\ \cmidrule(r){3-5} \cmidrule(r){7-9} \cmidrule(r){11-13} \cmidrule(r){15-17}
Network      & \multicolumn{1}{c|}{scanpath}                   & \multicolumn{1}{c}{CLE} & \multicolumn{1}{c}{GenLSTM} & \multicolumn{1}{c|}{Our} & \multicolumn{1}{c|}{scanpath} & \multicolumn{1}{c}{CLE} & \multicolumn{1}{c}{GenLSTM} & \multicolumn{1}{c|}{Our} & \multicolumn{1}{c|}{scanpath}                   & \multicolumn{1}{c}{CLE} & \multicolumn{1}{c}{GenLSTM} & \multicolumn{1}{c|}{Our} & \multicolumn{1}{c|}{scanpath} & \multicolumn{1}{c}{CLE} & \multicolumn{1}{c}{GenLSTM} & \multicolumn{1}{c}{Our} \\ \cmidrule(r){1-1} \cmidrule(r){2-2} \cmidrule(r){3-3} \cmidrule(r){4-4} \cmidrule(r){5-5} \cmidrule(r){6-6} \cmidrule(r){7-7} \cmidrule(r){8-8} \cmidrule(r){9-9} \cmidrule(r){10-10} \cmidrule(r){11-11} \cmidrule(r){12-12} \cmidrule(r){13-13} \cmidrule(r){14-14} \cmidrule(r){15-15} \cmidrule(r){16-16} \cmidrule(r){17-17}
ResNet-18    & \multicolumn{1}{r}{0.7991} & 0.7959                  & \textbf{0.8058}                  & 0.8003                  & 0.4539                       & 0.4557                  & \textbf{0.4750}                  & 0.4573                  & \multicolumn{1}{r}{0.7195} & 0.6445                  & 0.7383                  & \textbf{0.7444}                  & 0.3605                       & 0.2682                  & 0.3924                  & \textbf{0.4022}                  \\
ResNet-50    & \multicolumn{1}{r}{0.7928} & 0.8001                  & 0.7990                  & \textbf{0.8045}                  & 0.4598                       & 0.4633                  & 0.4636                  & \textbf{0.4717}                  & \multicolumn{1}{r}{0.6637} & 0.6236                  & 0.6810                  & \textbf{0.6898}                  & 0.2998                       & 0.2500                  & 0.3100                  & \textbf{0.3283}                  \\
DenseNet-121 & \multicolumn{1}{r}{0.8028} & \textbf{0.8057}                  & 0.7976                  & 0.8012                  & 0.4615                       & \textbf{0.4704}                  & 0.4539                  & 0.4626                  & \multicolumn{1}{r}{0.6356} & \textbf{0.7138}                  & 0.6953                  & 0.6687                  & 0.2615                       & 0.3342                  & \textbf{0.3363}                  & 0.3126                  \\
DenseNet-201 & \multicolumn{1}{r}{0.8061} & 0.8015                  & 0.8038                  & \textbf{0.8099}                  & 0.4671                       & 0.4659                  & 0.4648                  & \textbf{0.4767}                  & \multicolumn{1}{r}{0.6998} & 0.6916                  & 0.6954                  & \textbf{0.7039}                  & 0.3210                       & 0.2901                  & 0.3300                  & \textbf{0.3581}                  \\ \cmidrule(r){1-17}
\end{tabular}}
\end{table*}

\subsection{Performance of Scanpath based Classification}
\label{subsec:cfy_perform}
Our goal here is to evaluate the performance improvement achieved by using artificial visual scanpaths in our proposed model while performing multi-class multi-label disease classification on CXR images. As mentioned in Section~\ref{ExperimentalSettings}, we consider ResNet~\cite{he2016deep} and DenseNet~\cite{huang2017densely} architectures as the backbone CNN feature extractor in our model (See Fig.~\ref{fig: xray_class}). We consider  ResNet-18 and ResNet-50 in ResNet architecture and DenseNet-121 and DenseNet-201 in DenseNet architecture as the backbone CNN in our experiments. We evaluate our model which is trained on the training set ($MIMIC\_Train$) of MIMIC dataset for multi-label classification of CXR images into 14 diseases. We consider the widely used AUROC and AUPRC~\cite{davis2006relationship} as the evaluation measures. AUPRC has particularly been used for the quantitative evaluation of computer-aided systems that use X-ray imaging~\cite{balaram2022consistency, huang2020rectifying, 9718305}. To evaluate the benefit of using visual scanpaths for multi-label disease classification in CXR images, we present the results with and without using scanpaths in our model of Section~\ref{subsec:scan_fusion}. 

We perform within-dataset testing of our trained models using the test set $MIMIC\_Test$ of MIMIC dataset, and present the classification results in terms of average AUROC and AUPRC over all the images and the 14 classes in Table~\ref{tab:T2}. For cross-dataset testing of our model, we consider the complete CheXpert training dataset and show the classification results in terms of average AUROC and AUPRC over the 14 classes in Table~\ref{tab:T2}.  Table~\ref{tab:T2} shows that the scanpath guidance through our framework results in a considerable classification performance improvement (\% imprnt) in terms of average AUROC and AUPRC in all the cases of ResNet and DenseNet architecture except within-dataset results of DenseNet-121 where AUPRC disagrees with AUROC. The results also indicate that the performance boost is higher in ResNet-50 when compared to ResNet-18, both in the within-dataset and cross-dataset cases. Our observations here indicate that scanpath-based disease classification in CXR  using our approach not only performs better within-dataset but also generalizes better cross-dataset. This capability of the proposed approach is attributed to the guidance provided by the artificial scanpaths. 

\par Furthermore, in Table~\ref{tab:T3}, we have shown class-wise average AUROC and AUPRC scores over all the images for ResNet-50 backbone architecture.  As evident from the Table, the scanpath guidance provides a considerable performance improvement in terms of both the average AUROC and AUPRC scores in almost all the disease categories with very few exceptions. \textcolor{black}{A couple of exceptions are where the scanpath guidance does not help in cross-dataset adaptation for the Atelectasis and Support Devices classes, while it does help in the other $12$ classes. Such unfavourable observations during cross-dataset evaluation are usually because the patterns associated with the classes differ significantly between the two datasets.}


\par \textcolor{black}{In Table~\ref{tab:T4}, we compare the performance of our proposed scanpath based classification network when the scanpaths generated from the three different scanpath prediction approaches including ours are used. All these results are also compared to the results when scanpath is not used for guidance in the classification network in order to infer improvement due to the scanpath based guidance. As can be seen, the classification guidance using our scanpath prediction model results in the best improved performance in a majority of the cases. On the other hand, while guidance using GenLSTM results in performance close to ours, guidance by CLE improves the classification performance only in a few cases. Overall, the performances of the three models providing scanpath based guidance to the proposed classification model are in agreement with their scanpath prediction performances in Table~\ref{tab:T2}.}

\subsection{Ablation Study}
We perform an ablation study of our proposed scanpath-guided thoracic disease classifier to showcase the contribution of each part. Table~\ref{tab:ablation} shows four different baselines along with our proposed approach. We use AUROC and AUPRC as performance evaluation measures and ResNet-18 architecture as the backbone. All the models are trained on the MIMIC dataset and we use the whole training set of CheXpert for testing.
Ablation-1 is the basic ResNet-18 classifier without using scanpaths. In the proposed method, we remove the last linear layer in ResNet-18 to extract $N\times8\times8$ dimensional deepest spatial features from an image of size $256\times256$, where $N$ is the number of features. In our approach, we upscale the features to $N\times16\times16$ for scanpath-guided feature pooling with attention. Ablation-2 is our proposed approach without the attention module. In Ablation-3, instead of upscaling the features, we apply scanpath-guided feature pooling on the $N\times8\times8$ feature block itself. In Ablation-4, instead of upscaling the features, we remove more layers of the backbone architecture, and work on intermediate features with $16\times16$ spatial dimension. We can observe that attention to sequential features plays a crucial role as we compare Ablation-2 and the proposed approach in Table~\ref{tab:ablation}. Abstract features from the deeper layer are important for the classification, which can be seen when we compare results between Ablation-4 with the proposed approach. Scanpath attention in $8\times8$ feature space is not sufficient to achieve good performance as compared to $16\times16$ feature space, which can be seen by comparing Ablation-3 and the proposed approach.

\begin{table}[!tp]
	\centering
   \caption{An ablation study on our proposed scanpath-guided disease classification in CXR images.}
  \label{tab:ablation}
\scalebox{0.85}{
\begin{tabular}{l@{\hskip 0.12cm}|c@{\hskip 0.12cm}|c@{\hskip 0.12cm}|c@{\hskip 0.12cm}|c@{\hskip 0.12cm}|l@{\hskip 0.12cm}|l}
	\cmidrule(lr){1-7}
\small Modules & ResNet-18 & \begin{tabular}[c]{@{}c@{}c@{}}Intermediate \\ Features   \end{tabular}  & \begin{tabular}[c]{@{}c@{}c@{}}Deepest\\ Features \end{tabular} &  \begin{tabular}[c]{@{}c@{}}Attention \\  module \end{tabular}  &AUROC &AUPRC  \\ \cmidrule(lr){1-1} \cmidrule(lr){2-2} \cmidrule(lr){3-3} \cmidrule(lr){4-4} \cmidrule(lr){5-5} \cmidrule(lr){6-6} \cmidrule(lr){7-7}
Ablation-1&$\checkmark$&$\times$&$\times$&$\times$&0.7195&0.3605\\ & & & & & & \\
Ablation-2&$\checkmark$&$\times$&\begin{tabular}[c]{@{}c@{}}Upscaled \\ (16$\times$ 16) \end{tabular} &$\times$&0.7058&0.3323\\ & & & & & & \\
Ablation-3&$\checkmark$&$\times$&$\checkmark$(8$\times$ 8)&$\checkmark$&0.7389&0.3886\\ & & & & & & \\ 
Ablation-4&$\checkmark$&$\checkmark$(16$\times$ 16)&$\times$&$\checkmark$ &0.7131&0.3572\\ \cmidrule(lr){1-1} \cmidrule(lr){2-2} \cmidrule(lr){3-3} \cmidrule(lr){4-4} \cmidrule(lr){5-5} \cmidrule(lr){6-6} \cmidrule(lr){7-7}
\begin{tabular}[l]{@{}l@{}}Proposed \\  Method \end{tabular} &$\checkmark$&$\times$&\begin{tabular}[c]{@{}c@{}}Upscaled \\ (16$\times$ 16) \end{tabular}&$\checkmark$ &\textbf{0.7444}&\textbf{0.4022}\\\cmidrule(lr){1-7}
	\end{tabular}}

\end{table}

\section{Conclusion}
\label{sec:conclusion}

This paper introduces a novel approach to adopt the viewing patterns of radiologists while performing multi-label disease classification in Chest X-Ray (CXR) images. Due to the limited availability of recorded visual scanpaths on CXR images, we present an iterative scanpath predictor to generate artificial visual scanpaths on CXR images. A multi-class multi-label disease classification framework for CXR images is then proposed that works on a generated scanpath along with the input image features. Evaluation of the proposed visual scanpath predictor shows its capability to generate human-like scanpaths. Experiments related to multi-class multi-label disease classification on CXR images show that the use of generated visual scanpath results in performance improvement both in within and cross-dataset evaluations. Our work is the first step towards scanpath-based automated screening for diagnosis from CXR images and beyond.

\section{Aknowledgement}
Debashis Sen acknowledges the Science and Engineering Research Board (SERB), India for its assistance.

\end{document}